\documentclass{cpcauth}

\input{psfig}

\newcommand{\Z}{{\sf Z \!\!\! Z}}

\newcommand{\Sign}{\mbox{Sign}}
\newcommand{\up}{\uparrow}
\newcommand{\down}{\downarrow}

\usepackage{epsfig}

\usepackage{amssymb}

\begin{document}

\begin{frontmatter}

\title{From Spin Ladders to the 2-d $O(3)$ Model at Non-Zero Density}

\author{S. Chandrasekharan$^a$, B. Scarlet$^b$ and U.-J. Wiese$^{b,c}$}

\address{$^a$Department of Physics, Duke University, 
Box 90305, Durham, North Carolina 27708 \\
$^b$Center for Theoretical Physics, Laboratory for Nuclear Science and 
Department of Physics, \\
Massachusetts Institute of Technology, Cambridge, Massachusetts 02139 \\
$^c$Institute for Theoretical Physics, Bern University, 
Sidlerstrasse 5, 3012 Bern, Switzerland}

\begin{abstract}

The numerical simulation of various field theories at non-zero chemical 
potential suffers from severe complex action problems. In particular, QCD at 
non-zero quark density can presently not be simulated for that reason. A 
similar complex action problem arises in the 2-d $O(3)$ model --- a toy model 
for QCD. Here we construct the 2-d $O(3)$ model at non-zero density via
dimensional reduction of an antiferromagnetic quantum spin ladder in a magnetic
field. The complex action problem of the 2-d $O(3)$ model manifests itself as a
sign problem of the ladder system. This sign problem is solved completely with 
a meron-cluster algorithm.

\end{abstract}

\end{frontmatter}
 
\section{Introduction}

Numerical simulations of numerous quantum systems suffer from notorious sign
and complex action problems. For such systems the Boltzmann factor of a 
configuration in the path integral is in general complex and can hence not be 
interpreted as a probability. When the complex phase of the Boltzmann factor is
incorporated in measured observables, the fluctuations in the phase give rise 
to dramatic cancellations. In particular, for large systems at low temperatures
this leads to relative statistical errors that are exponentially large in both 
the volume and the inverse temperature. As a consequence, it is impossible in 
practice to study such systems with standard importance sampling Monte Carlo 
methods. 

Recently, some severe sign problems have been solved with meron-cluster 
algorithms \cite{Bie95,Cha99,Cha00,Cox99,Cha01}. Meron-clusters are used to 
identify canceling pairs of configurations with the same weight but opposite 
signs. Configurations with merons exactly cancel in the path integral. The 
Monte Carlo simulation can thus be restricted to the zero-meron sector with 
positive weights for which standard importance sampling works efficiently. 

It is natural to ask if the meron-concept can be applied to the complex action
problem in dense QCD. This is indeed the case in the limit of infinitely heavy
quarks in the Potts model approximation to QCD \cite {Alf01}. For dynamical
light quarks, on the other hand, it is not obvious if the meron concept can be 
applied to QCD. This seems most likely in the D-theory formulation of field 
theory in which 4-d QCD arises via dimensional reduction from a $(4+1)$-d
quantum link model \cite{Cha97,Bro99}. Physical gluons emerge as collective 
excitations of discrete variables \cite{Cha97} --- so-called quantum links --- 
which are gauge covariant generalizations of quantum spins. In this paper we 
study the D-theory formulation of the 2-d $O(3)$ model --- a toy model for 
QCD --- which arises via dimensional reduction from an antiferromagnetic spin 
ladder. In this case, the discrete variables are ordinary quantum spins. At the
level of the quantum spin system, the chemical potential of the $O(3)$ model 
manifests itself as an external magnetic field.

\section{Spin Ladders in a Magnetic Field and the 2-d $O(3)$ Model at Non-Zero
Density} 

We consider a ladder system of quantum spins $1/2$ on a square lattice of size 
$L \times L'$ ($L'$ even, $L \gg L'$) with periodic boundary conditions in both
directions. The spins located at the sites $x$ are described by operators 
${\bf S}_x$ with the usual commutation relations
\begin{equation}
[S_x^i,S_y^j] = i \delta_{xy} \epsilon_{ijk} S_x^k.
\end{equation}
The antiferromagnetic Hamilton operator ($J > 0$) 
\begin{equation}
\label{Hamiltonian}
H = J \sum_{x,i} {\bf S}_x \cdot {\bf S}_{x+\hat i} - 
{\bf B} \cdot \sum_x {\bf S}_x,
\end{equation}
couples the spins at the lattice sites $x$ and $x+\hat i$, where $\hat i$ is a 
unit-vector in the $i$-direction.

Chakravarty, Halperin and Nelson used a $(2+1)$-d effective field theory to
describe the low-energy dynamics of spatially 2-d quantum antiferromagnets
\cite{Cha88}. Chakravarty has applied this theory to quantum spin ladders with 
a large even number of coupled spin $1/2$ chains \cite{Cha96}. These systems 
are described by the effective action
\begin{eqnarray}
S[{\bf e}]&=&\int_0^\beta dt \int_0^L dx \int_0^{L'} dy \
\frac{\rho_s}{2}[\partial_x {\bf e} \cdot \partial_x {\bf e} 
+ \partial_y {\bf e} \cdot \partial_y {\bf e} \nonumber \\
&+&\frac{1}{c^2} \partial_t {\bf e} \cdot \partial_t {\bf e}],
\end{eqnarray}
where $\rho_s$ is the spin stiffness and $c$ is the spinwave velocity.

When the spin ladder is placed in a uniform external magnetic field ${\bf B}$,
the field couples to a conserved quantity --- the total spin. Hence, on the 
level of the effective theory, the magnetic field plays the role of a chemical 
potential, i.e. it appears as the time-component of an imaginary constant 
vector potential. As a consequence, the ordinary derivative 
$\partial_t {\bf e}$ is replaced by the covariant derivative $\partial_t 
{\bf e} + i {\bf B} \times {\bf e}$ and the action takes the form
\begin{eqnarray}
S[{\bf e}]&=&\int_0^\beta dt \int_0^L dx \int_0^{L'} dy \ 
\frac{\rho_s}{2}[\partial_x {\bf e} \cdot \partial_x {\bf e} 
+ \partial_y {\bf e} \cdot \partial_y {\bf e} \nonumber \\
&+&\frac{1}{c^2} (\partial_t {\bf e} + i {\bf B} \times {\bf e}) \cdot
(\partial_t {\bf e} + i {\bf B} \times {\bf e})].
\end{eqnarray}

For a sufficiently large number of coupled chains ($L' \gg c/\rho_s$) the 
system undergoes dimensional reduction to the 2-d $O(3)$ model with the action
\begin{eqnarray}
S[{\bf e}]&=&\int_0^\beta dt \int_0^L dx \
\frac{\rho_s L'}{2}[\partial_x {\bf e} \cdot \partial_x {\bf e} \nonumber \\
&+&\frac{1}{c^2} (\partial_t {\bf e} + i {\bf B} \times {\bf e}) \cdot
(\partial_t {\bf e} + i {\bf B} \times {\bf e})] \nonumber \\
&=&\int_0^{\beta c} d(ct) \int_0^L dx \
\frac{1}{2g^2}[\partial_x {\bf e} \cdot \partial_x {\bf e} \nonumber \\
&+&(\partial_{ct} {\bf e} + i {\bf \mu} \times {\bf e}) \cdot
(\partial_{ct} {\bf e} + i {\bf \mu} \times {\bf e})].
\end{eqnarray}
The effective coupling constant is given by $1/g^2 = \rho_s L'/c$ and the 
magnetic field appears as a chemical potential of magnitude $\mu = B/c$.

\section{Path Integral for Quantum Magnets}

To derive a path integral representation of the partition function we decompose
the Hamilton operator of eq.(\ref{Hamiltonian}) into five terms
\begin{equation}
H = H_1 + H_2 + ... + H_5.
\end{equation}
The various terms take the form
\begin{equation}
H_i = \!\!\! \sum_{\stackrel{x = (x_1,x_2)}{x_i even}} \!\!\! h_{x,i},
\, 
H_{i+2} = \!\!\! \sum_{\stackrel{x = (x_1,x_2)}{x_i odd}} \!\!\! 
h_{x,i},
\end{equation}
with $h_{x,i} = J {\bf S}_x \cdot {\bf S}_{x+\hat i}$ and 
\begin{equation}
H_5 = \!\!\! \sum_{\stackrel{x = (x_1,x_2)}{}} \!\!\! h_x,
\end{equation}
with $h_x = - B S_x^1$. The individual contributions to a given $H_i$ commute 
with each other, but two different $H_i$ do not commute. Using the 
Trotter-Suzuki formula we express the partition function as
\begin{eqnarray}
Z&=&\mbox{Tr} [\exp(- \beta H)] = 
\lim_{M \rightarrow \infty} \!\!\! \mbox{Tr} 
[\exp(- \epsilon H_1) \nonumber \\
&\times&\exp(- \epsilon H_2) ... \exp(- \epsilon H_5)]^M.
\end{eqnarray}
We have introduced $5 M$ Euclidean time slices with $\epsilon = \beta/M$ 
as the lattice spacing in the Euclidean time direction. We insert complete 
sets of eigenstates $|\up\rangle$ and $|\down\rangle$ with eigenvalues 
$S_x^3 = \pm 1/2$ between the factors $\exp(- \epsilon H_i)$.

The partition function is now expressed as a path integral
\begin{equation}
Z = \sum_s \Sign[s] \exp(- S[s]),
\end{equation}
over configurations of spins $s(x,t) = \up, \down$ on a $(2+1)$-dimensional 
space-time lattice of points $(x,t)$. The Boltzmann factor $\exp(- S[s])$
is a product of space-time plaquette contributions
$\exp\{- S[s(x,t),s(y,t),s(x,t+1),s(y,t+1)]\}$ with
\begin{eqnarray}
\label{Boltzmann1}
\exp(- S[\up,\up,\up,\up])&=&\exp(- S[\down,\down,\down,\down]) = \nonumber \\ 
&=&\exp(- \epsilon J/2), \nonumber \\
\exp(- S[\up,\down,\up,\down])&=&\exp(- S[\down,\up,\down,\up]) = \nonumber \\ 
&=&\cosh(\epsilon J/2), \nonumber \\
\exp(- S[\up,\down,\down,\up])&=&\exp(- S[\down,\up,\up,\down]) = \nonumber \\
&=&\sinh(\epsilon J/2),
\end{eqnarray}
as well as the time-like bond contributions $\exp\{- S[s(x,t),s(x,t+1)]\}$ with
\begin{eqnarray}
\label{Boltzmann2}
&&\exp(- S[\up,\up]) = \exp(- S[\down,\down]) = \cosh(\epsilon B/2),
\nonumber \\
&&\exp(- S[\up,\down]) = \exp(- S[\down,\up]) = \sinh(\epsilon B/2).
\end{eqnarray}

The sign of a configuration, $\Sign[s]$, also is a product of space-time 
plaquette contributions $\mbox{Sign}[s(x,t),s(y,t),s(x,t+1),s(y,t+1)]$ with
\begin{eqnarray}
&&\mbox{Sign}[\up,\up,\up,\up]) = \mbox{Sign}[\down,\down,\down,\down]) = 1,
\nonumber \\ 
&&\mbox{Sign}[\up,\down,\up,\down]) = 
\mbox{Sign}[\down,\up,\down,\up]) = 1, \nonumber \\
&&\mbox{Sign}[\up,\down,\down,\up]) = \mbox{Sign}[\down,\up,\up,\down]) = 
- 1.
\end{eqnarray}
Figure 1 shows two spin configurations in $(1+1)$ dimensions. The first
configuration is completely antiferromagnetically ordered and has 
$\Sign[s] = 1$. The second configuration contains one interaction plaquette
with configuration $[\down,\up,\up,\down]$ which contributes 
$\Sign[\down,\up,\up,\down] = - 1$, such that the whole configuration has 
$\Sign[s] = - 1$. 

The central observable of our study is the uniform magnetization 
${\bf M} = \sum_x {\bf S}_x$.
\begin{figure}[htb]
\vbox{
\epsfig{file=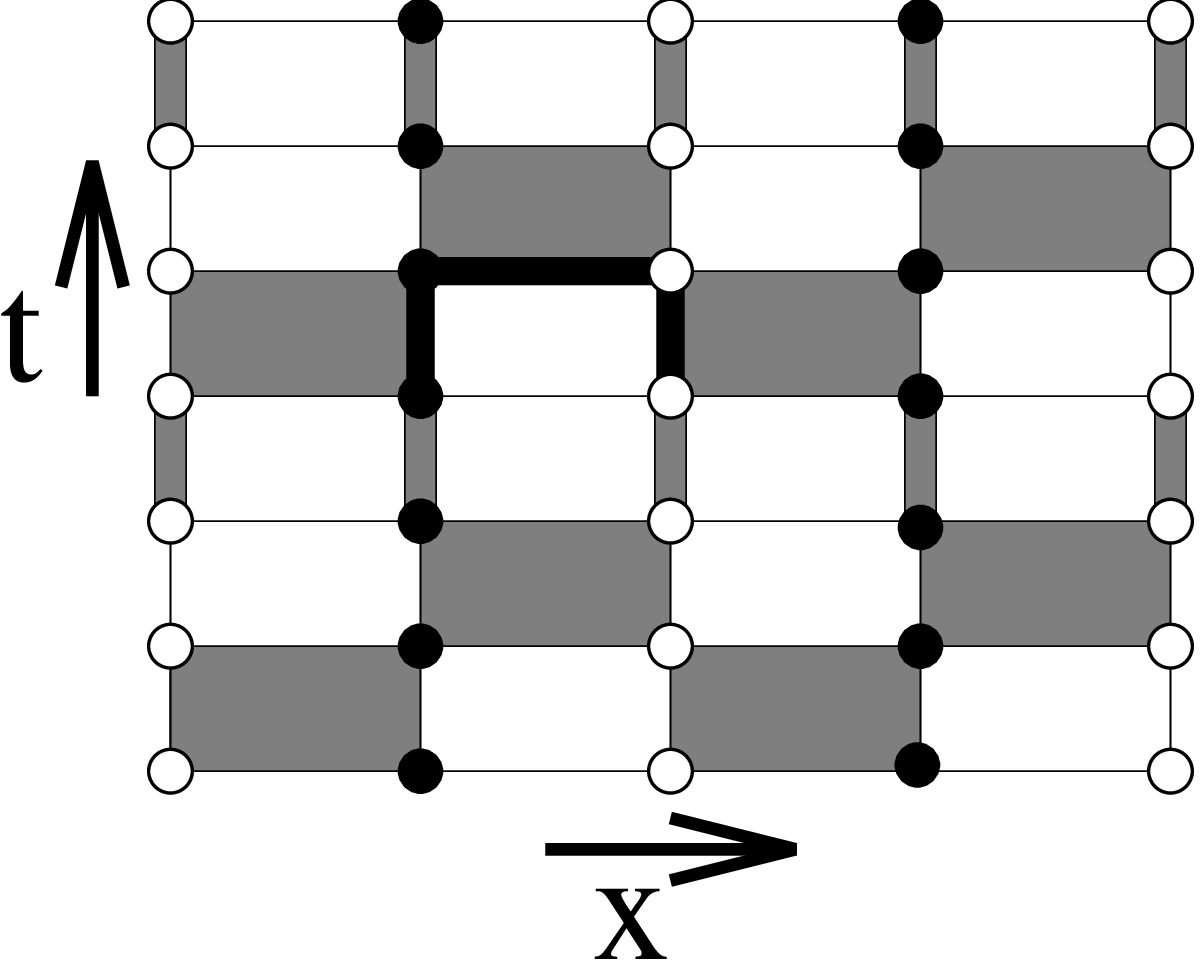,
width=4.5cm,angle=270,
bbllx=0,bblly=0,bburx=225,bbury=337}}
\vspace{1.1cm}
\vbox{
\hspace{0.8cm}
\epsfig{file=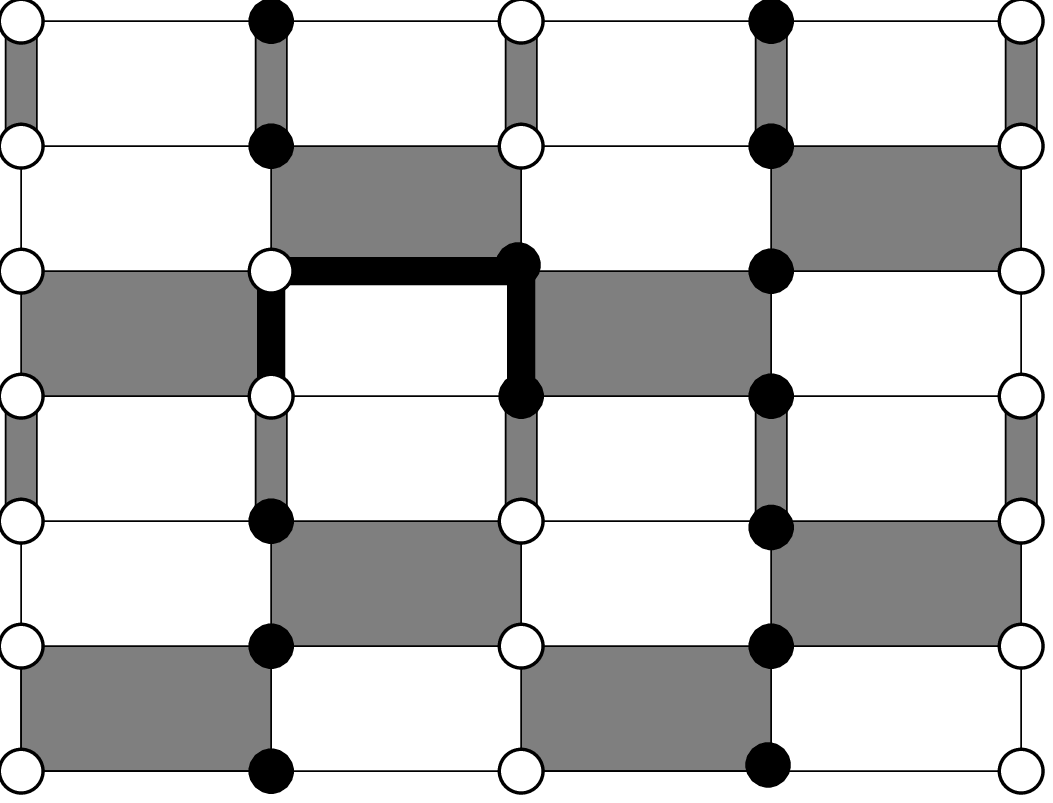,
width=4.5cm,angle=270,
bbllx=0,bblly=0,bburx=225,bbury=337}}
\caption{\it Two spin configurations in $(1+1)$ dimensions. The shaded
plaquettes and time-like bonds carry the interaction. Filled dots represent 
spin up and open circles represent spin down. The first configuration has
$\Sign[s] = 1$ and the second configuration has $\Sign[s] = - 1$. The fat black
line represents a meron-cluster. The other clusters are not shown. Flipping the
meron-cluster changes one configuration into the other and changes $\Sign[s]$.}
\end{figure}

\section{Meron-Cluster Algorithm}

The meron-cluster algorithm is based on a cluster algorithm for a modified 
model without the sign factor. Quantum spin systems without a sign problem can 
be simulated very efficiently with the loop-cluster algorithm 
\cite{Eve93,Wie94,Eve97,Bea96}. The idea behind the algorithm is to decompose a
configuration into clusters which can be flipped independently. Each lattice 
site belongs to exactly one
cluster. When the cluster is flipped, the spins at all the sites on the cluster
are changed from up to down and vice versa. The decomposition of the lattice
into clusters results from connecting neighboring sites on each space-time 
interaction plaquette or time-like interaction bond according to probabilistic 
cluster rules. A set of connected sites defines a cluster. In this case the 
clusters are open or closed strings. The cluster rules are constructed so as to
obey detailed balance. To show this property we first write the space-time 
plaquette 
Boltzmann factors as
\begin{eqnarray}
\label{cluster1}
&&\exp(- S[s(x,t),s(y,t),s(x,t+1),s(y,t+1)]) = \nonumber \\
&&A \delta_{s(x,t),s(x,t+1)} \delta_{s(y,t),s(y,t+1)} + \nonumber \\
&&B \delta_{s(x,t),-s(y,t)} \delta_{s(x,t+1),-s(y,t+1)}.
\end{eqnarray}
The $\delta$-functions specify which sites are connected and thus belong to the
same cluster. The coefficients $A$ and $B$ determine the relative probabilities
for different cluster break-ups of an interaction plaquette. For example, $A$ 
determines the probability with which sites are connected with their time-like 
neighbors and $B$ determines the probability for connections with space-like 
neighbors. Inserting the expressions from eq.(\ref{Boltzmann1}) one finds
\begin{eqnarray}
\label{balance1}
\exp(- S[\up,\up,\up,\up])&=&\exp(- S[\down,\down,\down,\down]) = \nonumber \\ 
&=&\exp(- \epsilon J/2) = A, \nonumber \\
\exp(- S[\up,\down,\up,\down])&=&\exp(- S[\down,\up,\down,\up]) = \nonumber \\ 
&=&\cosh(\epsilon J/2) = A + B, \nonumber \\
\exp(- S[\up,\down,\down,\up])&=&\exp(- S[\down,\up,\up,\down]) = \nonumber \\ 
&=&\sinh(\epsilon J/2) = B.
\end{eqnarray}
Similarly, the time-like bond Boltzmann factors are expressed as
\begin{eqnarray}
\label{cluster2}
&&\exp(- S[s(x,t),s(x,t+1)]) = \nonumber \\
&&C \delta_{s(x,t),s(x,t+1)} + D.
\end{eqnarray}
The probability to connect spins with their time-like neighbors is $C/(C+D)$.
The spins remain disconnected with probability $D/(C+D)$. Inserting the 
expressions from eq.(\ref{Boltzmann2}) one obtains
\begin{eqnarray}
\label{balance2}
\exp(- S[\up,\up])&=&\exp(- S[\down,\down]) = \nonumber \\
&=&\cosh(\epsilon B/2) = C + D, \nonumber \\
\exp(- S[\up,\down])&=&\exp(- S[\down,\up]) = \nonumber \\
&=&\sinh(\epsilon B/2) = D.
\end{eqnarray}
The cluster rules are illustrated in table 1.
\begin{table}[htb]
\catcode`?=\active \def?{\kern\digitwidth}
  \begin{center}
    \leavevmode
    \begin{tabular}{|c|c|}
	\hline
	configuration&break-ups\\
	\hline\hline
\hspace{0.4cm}
        \begin{minipage}[c]{2.5cm}
		\epsfig{file=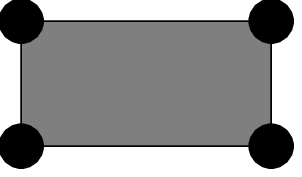,width=1.5cm,angle=270,
		bbllx=-5,bblly=0,bburx=55,bbury=85}
	\end{minipage} &
\hspace{0.4cm}
        \begin{minipage}[c]{2.5cm}
		\vbox{\begin{center}
		\epsfig{file=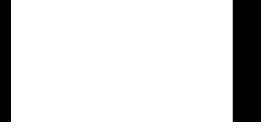,width=1.5cm,angle=270,
		bbllx=-20,bblly=0,bburx=55,bbury=85} 

		\hspace{-0.2cm}\vspace{0.1cm}A
		\end{center}}
	\end{minipage}
	\\
	\hline
\hspace{0.4cm}
        \begin{minipage}[c]{2.5cm}
		\epsfig{file=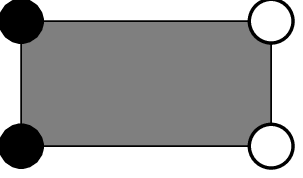,width=1.5cm,angle=270,
		bbllx=-5,bblly=0,bburx=55,bbury=85}
	\end{minipage} & $\begin{array}{c}
\hspace{0.4cm}
        \begin{minipage}[c]{2.5cm}
		\vbox{\begin{center}
		\epsfig{file=table_1_2.eps,width=1.5cm,angle=270,
		bbllx=-20,bblly=0,bburx=55,bbury=85} 

        	\hspace{-0.2cm}\vspace{0.1cm}A
		\end{center}}
	\end{minipage}
        \\
\hspace{0.4cm}
        \begin{minipage}[c]{2.5cm}
		\vbox{\begin{center}
		\epsfig{file=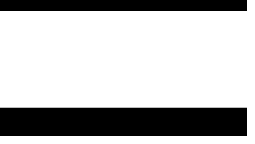,width=1.5cm,angle=270,
		bbllx=-20,bblly=0,bburx=55,bbury=85} 

                \hspace{-0.2cm}\vspace{0.1cm}B
	\end{center}}
	\end{minipage} \end{array}$
	\\
	\hline
\hspace{0.4cm}
        \begin{minipage}[c]{2.5cm}
		\epsfig{file=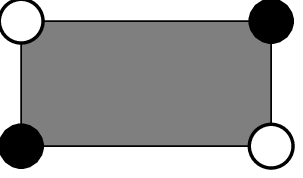,width=1.5cm,angle=270,
		bbllx=-5,bblly=0,bburx=55,bbury=85}
	\end{minipage} &
\hspace{0.4cm}
        \begin{minipage}[c]{2.5cm}
		\vbox{\begin{center}
		\epsfig{file=table_3_2.eps,width=1.5cm,angle=270,
		bbllx=-20,bblly=0,bburx=55,bbury=85} 

        	\hspace{-0.2cm}\vspace{0.1cm}B
		\end{center}}
	\end{minipage}
	\\
	\hline   
\hspace{1cm}
        \begin{minipage}[c]{1.5cm}
		\epsfig{file=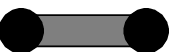,width=1.5cm,angle=270,
		bbllx=-5,bblly=0,bburx=55,bbury=85}
	\end{minipage} &
\hspace{0.4cm}
        \begin{minipage}[c]{0.5cm}
		\vbox{\begin{center}
		\epsfig{file=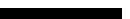,width=1.5cm,angle=270,
		bbllx=-10,bblly=-5,bburx=45,bbury=5}
 
        	\hspace{0cm}\vspace{0.1cm}C
		\end{center}}
	\end{minipage}
\hspace{2.4cm}
        \begin{minipage}[c]{0.5cm}
		\vbox{\begin{center}
		\epsfig{file=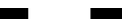,width=1.5cm,angle=270,
		bbllx=-10,bblly=-5,bburx=45,bbury=5} 
        	
                \hspace{0cm}\vspace{0.1cm}D
		\end{center}}
	\end{minipage}
	\\
	\hline  
\hspace{1.0cm}
        \begin{minipage}[c]{1.5cm}
		\epsfig{file=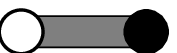,width=1.5cm,angle=270,
		bbllx=-5,bblly=0,bburx=55,bbury=85}
	\end{minipage} &
\hspace{0.4cm}
        \begin{minipage}[c]{0.5cm}
		\vbox{\begin{center}
		\epsfig{file=table_4_3.eps,width=1.5cm,angle=270,
		bbllx=-10,bblly=-5,bburx=45,bbury=5} 
        	
                \hspace{0cm}\vspace{0.1cm}D
		\end{center}}
	\end{minipage}
	\\
	\hline   
    \end{tabular}
\caption{\it Cluster break-ups of various plaquette and time-like bond 
configurations together with their relative probabilities $A,B,C,D$. Filled 
dots represent spin up, open circles represent spin down, and the fat lines are
the cluster connections.}
\end{center}
\end{table}

Eqs.(\ref{cluster1},\ref{cluster2}) can be viewed as a representation of the 
original model as a random cluster model. The cluster algorithm operates in two
steps. First, a cluster break-up is chosen for each space-time interaction 
plaquette or time-like interaction bond according to the above probabilities. 
This effectively replaces the original Boltzmann weight of a configuration with
a set of constraints represented by the $\delta$-functions associated with the 
chosen break-ups. The constraints imply that the spins in one cluster can only
be flipped together. Second, every cluster is flipped with probability $1/2$. 
When a cluster is flipped, the spins on all sites that belong to the cluster 
are flipped from up to down and vice versa. Eqs.(\ref{balance1},\ref{balance2})
ensure that the cluster algorithm obeys detailed balance.

The above cluster rules were first used in a simulation of the Heisenberg 
antiferromagnet \cite{Wie94} in the absence of a magnetic field. In that case
there is no sign problem. Then the corresponding loop-cluster algorithm is 
extremely efficient and has almost no detectable autocorrelations. When a 
magnetic field is switched on the situation changes. When the magnetic field 
points in the direction of the spin quantization axis (the $3$-direction in our
case) there is no sign problem. However, the magnetic field then explicitly 
breaks the $\Z(2)$ flip symmetry on which the cluster algorithm is based, and 
the clusters can no longer be flipped with probability $1/2$. Instead the flip 
probability is determined by the value of the magnetic field and by the 
magnetization of the cluster. When the field is strong, flips of magnetized
clusters are rarely possible and the algorithm becomes inefficient. To avoid 
this, we have chosen the magnetic field to point in the $1$-direction, i.e. 
perpendicular to the spin quantization axis. In that case, the cluster flip 
symmetry is not affected by the magnetic field, and the clusters can still be 
flipped with probability $1/2$. However, one now faces a sign problem and the 
cluster algorithm becomes extremely inefficient again. Fortunately, using the 
meron concept the sign problem can be eliminated completely and the efficiency 
of the original cluster algorithm can be maintained even in the presence of a 
magnetic field.

\section{Meron-Clusters and the Sign Problem}

Let us consider the effect of a cluster flip on the sign. The flip of a 
meron-cluster changes $\Sign[s]$, while the flip of a non-meron-cluster leaves 
$\Sign[s]$ unchanged. An example of a meron-cluster is shown in figure 1. When 
the meron cluster is flipped, the first configuration with $\Sign[s] = 1$ turns
into the second configuration with $\Sign[s] = - 1$. This property of the 
cluster is independent of the orientation of any other cluster. Since flipping 
all spins leaves $\Sign[s]$ unchanged, the total number of meron-clusters is 
always even.

The meron concept allows us to gain an exponential factor in statistics. Since 
all clusters can be flipped independently with probability $1/2$, one can 
construct an improved estimator for $\langle \Sign \rangle$ by averaging 
analytically over the $2^{N_C}$ configurations obtained by flipping the $N_C$ 
clusters in a configuration in all possible ways. For configurations that 
contain merons, the average $\Sign[s]$ is zero because flipping a single 
meron-cluster leads to a cancellation of contributions $\pm 1$. Hence only the
configurations without merons contribute to $\langle \Sign \rangle$. The 
probability for having a configuration without merons is equal to 
$\langle \Sign \rangle$ and is exponentially suppressed with the space-time 
volume. The vast majority of configurations contains merons and contributes an 
exact $0$ to $\langle \Sign \rangle$ instead of a statistical average of 
contributions $\pm 1$. In this way the improved estimator leads to an 
exponential gain in statistics. One can show that the contributions from the
zero-meron sector are always positive. 

One can also find a simple expression for the improved estimator for the
magnetization $\langle M^1 \rangle$ in terms of a winding number $W_l$ of 
closed loops which result from joining open string clusters. One can define 
$W_l$ for each loop to be its temporal winding number. If a particular loop is 
not composed of open string clusters then $W_l = 0$. With this definition of 
$W_l$ it is easy to show that
\begin{equation}
\label{ratio}
\langle M^1 \rangle = 
\frac{\langle \delta_{N,0} \sum_l W_l \rangle}{2 \langle \delta_{N,0} \rangle},
\end{equation}
where $N$ is the number of meron-clusters.

Since the magnetization gets non-vanishing contributions only from the 
zero-meron sector, it is unnecessary to generate any configuration that 
contains meron-clusters. This observation is the key to the solution of the 
sign problem. In fact, one can gain an exponential factor in statistics by 
restricting the simulation to the zero-meron sector, which represents an 
exponentially small fraction of the whole configuration space. We visit all 
plaquette and time-like bond interactions one after the other and choose new 
pair connections between the sites according to the above cluster rules. A 
newly proposed pair connection that takes us out of the zero-meron sector is
rejected. After visiting all plaquette and time-like bond interactions, each 
cluster is flipped with probability $1/2$ which completes one update sweep. 
In practice, it is advantageous to occasionally generate configurations 
containing merons even though they do not contribute to our observable, because
this reduces the autocorrelation times. 

Figure 2 shows the magnetization density of antiferromagnetic quantum spin 
ladders with $L' = 4$ compared to analytic results (valid for large $L'$)
obtained with the Bethe ansatz \cite{Wie85} using the massgap $m = 0.141(2)/a$ 
\cite{Syl97} and the spinwave velocity $c = 1.657(2) Ja$ \cite{Bea98}. The 
agreement is remarkable and involves no adjustable free parameters.
\begin{figure}[htb]
\begin{center}
\epsfig{file=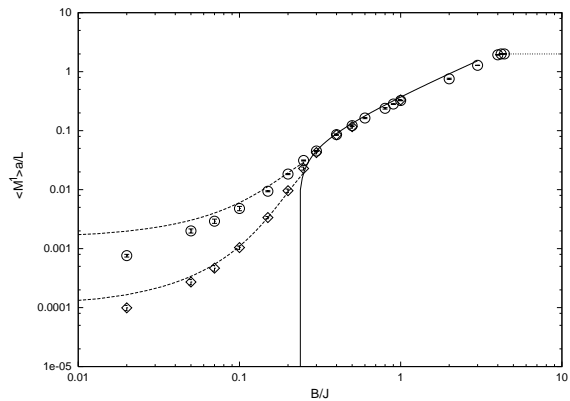,
width=5.3cm,angle=270,
bbllx=53,bblly=50,bburx=554,bbury=770}
\end{center}
\caption{\it Magnetization density $\langle M^1 \rangle/L$ of quantum spin 
ladders consisting of $L' = 4$ coupled chains as a function of the magnetic 
field $B$. The numerical data are for two systems: one of size $L = 20$ at 
inverse temperature $\beta J = 15$ (circles) and the other for $L = 40$ at 
$\beta J = 24$ (diamonds). The solid curve is the infinite volume, zero 
temperature analytic result, while the two dashed curves are finite volume, 
non-zero temperature analytic results for the two simulated systems in the 
intermediate $B$ region. The dotted curve represents saturation of the 
magnetization per spin at $1/2$.}
\end{figure}

\section{Conclusions}

Using D-theory, the 2-d $O(3)$ model at non-zero chemical potential has been
obtained from dimensional reduction of a $(2+1)$-d quantum spin ladder in a
magnetic field. The resulting sign problem has been solved completely with a
meron-cluster algorithm. This is the first time that this toy model for QCD has
been simulated efficiently at non-zero chemical potential. The next challenge 
is to address the complex action problem of dense QCD. In D-theory quarks arise
as domain wall fermions and gluons emerge as collective excitations of quantum 
links. Hence, one needs to generalize the meron-concept to quantum link models 
as well as to domain wall fermions.


\begin{thebibliography}{10}

\bibitem{Bie95} 
W. Bietenholz, A. Pochinsky and U.-J. Wiese, Phys. Rev. Lett. 75 (1995) 4524.

\bibitem{Cha99}
S. Chandrasekharan and U.-J. Wiese, Phys. Rev. Lett. 83 (1999) 3116.

\bibitem{Cha00}
S. Chandrasekharan, J. Cox, K. Holland and U.-J. Wiese, 
Nucl. Phys. B576 (2000) 481.

\bibitem{Cox99}
J. Cox, C. Gattringer, K. Holland, B. Scarlet and U.-J. Wiese, 
Nucl. Phys. (Proc. Suppl.) 83 (2000) 777.

\bibitem{Cha01}
S. Chandrasekharan and J. Osborn, cond-mat/0109424.

\bibitem{Alf01}
M. Alford, S. Chandrasekharan, J. Cox and U.-J. Wiese, 
Nucl. Phys. B602 (2001) 61.

\bibitem{Cha97}
S. Chandrasekharan and U.-J. Wiese, Nucl. Phys. B492 (1997) 455.

\bibitem{Bro99}
R. Brower, S. Chandrasekharan and U.-J. Wiese, Phys. Rev. D60 (1999) 094502.

\bibitem{Cha88}
S. Chakravarty, B. I. Halperin and D. R. Nelson, Phys. Rev. Lett. 60 (1988) 
1057; Phys. Rev. B39 (1989) 2344.

\bibitem{Cha96}
S. Chakravarty, Phys. Rev. Lett. 77 (1996) 4446.

\bibitem{Eve93}
H. G. Evertz, G. Lana and M. Marcu, Phys. Rev. Lett. 70 (1993) 875.

\bibitem{Wie94}
U.-J. Wiese and H.-P. Ying, Z. Phys. B93 (1994) 147.

\bibitem{Eve97}
H. G. Evertz, The loop algorithm, in Numerical Methods for Lattice Quantum
Many-Body Problems, ed. D. J. Scalapino, Addison-Wesley Longman, Frontiers in
Physics.

\bibitem{Bea96}
B. B. Beard and U.-J. Wiese, Phys. Rev. Lett. 77 (1996) 5130.

\bibitem{Wie85}
P. B. Wiegmann, Phys. Lett. B152 (1985) 209.

\bibitem{Syl97}
O. F. Syljuasen, S. Chakravarty and M. Greven, Phys. Rev. Lett. 78 (1997) 4115.

\bibitem{Bea98}
B. B. Beard, R. J. Birgeneau, M. Greven and U.-J. Wiese, Phys. Rev. Lett.
80 (1998) 1742.

\end{thebibliography}
\end{document}